\documentclass [12pt]{article}

\usepackage{graphicx}

\usepackage{amsfonts}

\usepackage{amscd}

\usepackage{authblk}

\newcommand{\be}{\begin{equation}}
\newcommand{\ee}{\end{equation}}
\newcommand{\bea}{\begin{eqnarray}}
\newcommand{\eea}{\end{eqnarray}}
\title{\Large  Images of  the lines under the MS transformations and the Concept of Velocity in the DSR theories }

\author{ Nosrtollah Jafari }

\affil{  \textit{Science and Research Branch, Islamic Azad University, Hesarak, Tehran, 14778, Iran.} \thanks{ nosrat.jafari@gmail.com }   }

\begin{document}

\maketitle

\begin{abstract}

The effect of the Maguejo-Smolin (MS) transformations on a straight line in the energy-
momentum space will be studied. We will interpret the slope of this line as velocity
$dE/dp$, which can leads to addition rule for the velocities in the MS
doubly special relativity (DSR) case. Relation between two expressions  $dE/dp$ and $p/E$ for
velocity in the momentum space will be investigated more and the energy dependency of
the velocities in the DSR theories is related to the geometrical properties of the lines under DSR
transformations. The images of two parallel lines under the MS transformations will be
studied and we will compute crossing point of these lines under the MS transformations
 in the energy-momentum space. The linear-fractional transformations don't keep parallelism.
 The crossing point is on a line in the energy-momentum space with a constant momentum $E_p/c$.

\end{abstract}

\vspace{2cm}

\noindent Keywords: Linear-Fractional Transformations, Doubly Special Relativity,
Geometry of the Energy-Momentum Space.

\newpage

\section{Introduction}

We begin by studying the effect of the Lorentz transformations in the (1 + 1)
dimensional energy-momentum space on a straight line. The Lorentz transformations map a
straight line to another straight line in the energy-momentum space. From
this property we can obtain the Einstein's addition rule for the velocities from slope of this line.
 In continuation, we study the effect of the Maguejo-Smolin (MS) transformations on a straight line
in the energy-momentum space. Also, we interpret the slope of this line
as a velocity $dE/dp$ and we
will obtain an addition rule for velocities in the MS doubly special
relativity case. However, introducing velocities in the doubly special relativity
(DSR) theories is a difficult problem, and there is a long discussion for
introducing a convenient velocity \cite{Ame1, Luk, Rem, Gra, Kos, Mig1, Das, Alo}  .
For example, in the MS DSR which is a very simple example of DSR theories,
introducing  $dE/dp$ as velocity will bring the properties of particles,
such as energy, momentum or even mass to the definition of the velocity. In this paper,
we relate the dependency of the velocity on the energy or momentum to the geometrical properties of
the MS transformations and image of a line under these transformations.

Images of the two parallel lines in the energy- momentum under the Lorentz transformations
will remain parallel, but under the MS transformations these lines will cross each other.
We will find the coordinates of this crossing point which depends on the energy scale of 
the MS DSR i.e. the Planck Energy $E_p$ .

Finally, we will compute a general formula for combination rule of the velocities in the MS DSR
by taking derivatives from the MS transformations. This computation will bring our arguments and
calculations to a harmony.

\newpage

\section{ Image of a line in the Energy-Momentum space }

\subsection{ The Lorentz Transformations case}

 The Lorentz transformations in the (1+1) dimensions are

   \be  \label{eq1} \left\{\begin{array}{cl}  p'_1=\gamma ( p_1 - v p_0),\\\\
   p'_0=\gamma ( p_0 - v p_1), \end{array}\right.\ee  here we have taken the $p_1$ component
   in $x$-direction and we have assumed $c=1$.

   We take the equation of a line in the momentum space which passing through $(\varphi_0, \varphi _1)$ point as
\be \label{eq 2} p_0 = \psi p_1 + \Pi_0, \ee
where we have put $\Pi_0 = \varphi_0 -\psi\varphi_1$. For massive particles this line can
be interpreted as a tangent line in the $(\varphi_0, \varphi_1)$ point to the mass-shell 
of a massive particle in the Energy-Momentum space. This tangent line has been shown in Fig. 1.
 In fact, the mass of this massive particle will be \be m^2 = \varphi_0^2 - \varphi_1^2. \ee

The Lorentz transformations Eq.(\ref{eq1}) map the line in Eq.(\ref{eq 2}) to
 \be p'_0 = \Big(\frac{ \psi - u}{1 - \psi u} \Big)p'_1+ \frac{\Pi_0}{\gamma(1 -u)}= \psi'p'_1 + \Pi'_0 . \ee

If we interpret $ \psi$ as the velocity of the particle, then the linearity of the
Lorentz transformations will show that $\psi'$ is in the form of the Einstein's
combination rule for velocities
\be  \label{eq4} \psi'= \frac{ \psi - u}{1 - \psi u}:= v_H. \ee
The interpretation of $\psi $ as the velocity is not so strange, because in classical
mechanics we have the Hamiltonian expression
\be  v_H =\frac{dE}{dp} \ee
for the velocity. But we have also another expression
 \be v_G =\frac{p}{E}  \ee which can be interpreted as the velocity.
 Our notations for these velocities is from \cite{Mig1}.

\begin{figure}
  \centering
    \includegraphics[width=1.5\textwidth]{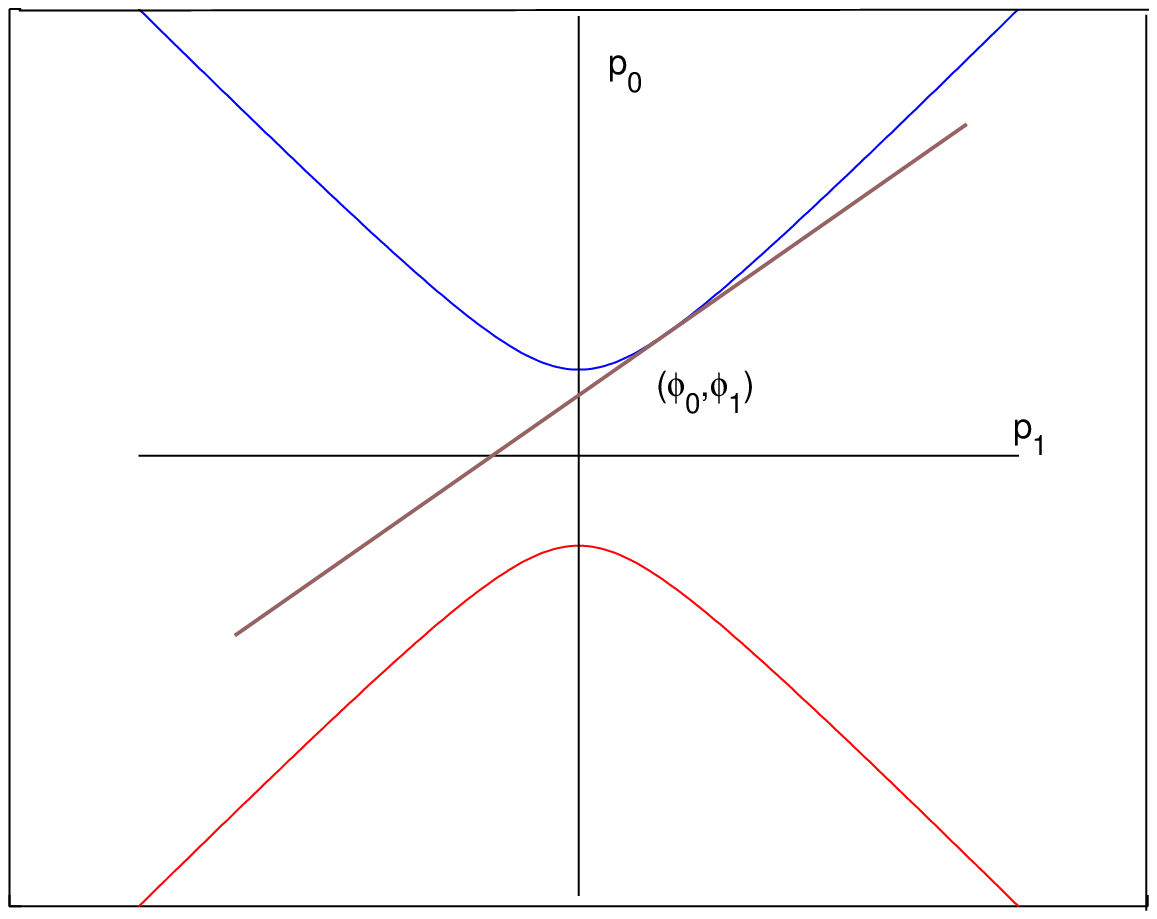}\\
  \caption{Tangent line to the mass shell at $(\varphi_0, \varphi_1)$  }\label{g1}
\end{figure}

\subsection{The MS Transformations case}

The MS transformations in (1+1) dimensions are
 \be  \left\{\begin{array}{cl} p'_0 =\frac {\gamma(p_0 - up_1)}{1 + l_p(\gamma - 1)p_0 - l_p \gamma u p_1 },\\\\
 p'_1 = \frac{\gamma (p_1 - up_0)}{1 + l_p(\gamma - 1)p_0 - l_p \gamma u p_1 }, \end{array}\right.\ee
%in which the $S′$ frame moves with constant $u$ with respect to $S$ frame and $l_p$
in which $l_p$ is the Planck length. These transformations map the straight line
\be p_0 = \psi p_1 + \Pi_0 \ee to

\be   p'_0= \Big[ \frac{\psi - u + \Pi_0 l_p u}{ 1- \psi u - \Pi_0 l_p (\gamma -1)/ \gamma }      \Big] p'_1 +
\frac{\Pi_0}{\Big[ \gamma-  \gamma \psi u - \Pi_0 l_p (\gamma -1) \Big] }.  \ee

The combinations rule for the velocities will be
\be \label{eq13}  v'_H=  \frac{v - u + \Pi_0 l_p u}{ 1- v u - \Pi_0 l_p (\gamma -1)/ \gamma }.  \ee

As in the Lorentz case we can interpret $ \psi$ as the velocity. Please, note that
this expression is a combination rule for a particle with constant velocity $v$
in the $S$ system. The more general formula will be obtained in the section 5.
Here, unusual matter is the appearance of energy and momentum
through $ \Pi_0=\varphi_0 - \psi\varphi_1 $ in the expression
for the relative velocity.

\section{Images of two parallel lines in the Energy-Momentum space}

\subsection{ The Lorentz transformations case}

Under the Lorentz transformations the images of two parallel lines are parallel lines.
However, the interpretation of a line in the Energy-Momentum space is difficult, but as
discussed in section 2, we can draw a straight line tangent to the every point of
the mass shell of a particle in the energy-momentum space. We can also draw two parallel lines
as tangent lines to the mass shell of a massive particle which are shown in the Fig. 2 . Images of these
parallel lines under the Lorentz transformations will be parallel, but rotated in the $(p_0,p_1)$
plane and we have not drawn them. However, we can assume any two parallel lines in the energy-momentum
space, the parallel lines which are tangent to mass shell in the Fig. 2 are good for study of the
effect of the Lorentz transformations on the rest mass of a particle and its conjugate antiparticle.

\begin{figure}
  \centering
    \includegraphics[width=1.5\textwidth]{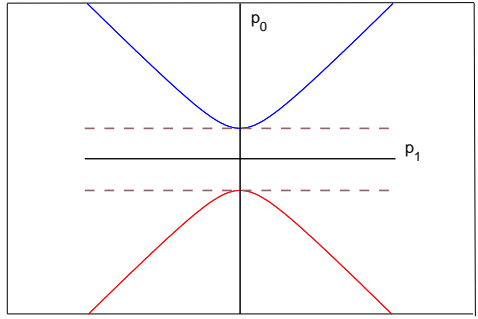}\\
  \caption{Two parallel lines in the Energy-momentum space which are tangent to the mass shell}\label{g1}
\end{figure}

\subsection{ The MS transformations case}

We want to study the effect of the  MS transformations on two parallel lines

\be  \label{eq} \left\{\begin{array}{cl}  p_0=\psi p_1 + \Pi_0 ,\\\\
   \tilde{p}_0=  \psi p_1 + \tilde{\Pi}_0\emph{} , \end{array}\right.\ee
in the energy-momentum space.
The MS transformations map these lines to
\be  \label{eq} \left\{\begin{array}{cl}  p'_0= \Big[ \frac{\psi - u + \Pi_0 l_p u}{ 1- \psi u - \Pi_0 l_p (\gamma -1)/ \gamma }
      \Big] p'_1 + \Pi_0/\Big[ \gamma-  \gamma \psi u - \Pi_0 l_p (\gamma -1) \Big],  \\\\
   \tilde{p}'_0= \Big[ \frac{\psi - u + \tilde{\Pi}_0 l_p u}{ 1- \psi u - \tilde{\Pi}_0 l_p (\gamma -1)/ \gamma }
      \Big] p'_1 + \tilde{\Pi}_0/\Big[ \gamma-  \gamma \psi u - \tilde{\Pi}_0 l_p (\gamma -1) \Big], \end{array}\right.\ee
These lines cross each other at
\be (p'_0,p'_1)= \Big( - \psi'  \frac{1-\psi u}{ l_p\Big[  u + \psi(1-\gamma)/\gamma   \Big]} + \Pi'_0,
 -\frac{1-\psi u}{l_p \Big[  u + \psi(1-\gamma)/\gamma   \Big]} \Big), \ee
in these formulas $\psi'$  and $\Pi'_0$  depend on the initial coordinates $ (\varphi_0, \varphi_1 )$ and are given by
\be  \psi' = \frac{v - u + \Pi_0 l_p u}{ 1- v u - \Pi_0 l_p (\gamma -1)/ \gamma }, \ee
and
\be \Pi'=\frac{\Pi_0}{\Big[ \gamma-  \gamma \psi u - \Pi_0 l_p (\gamma -1) \Big] } .\ee

For higher velocities $\gamma >> 1$, the expressions for the coordinates of the crossing point can be rewritten in a more simpler form
\bea  \label{eq 20}(p'_0, p'_1)&=& \Big( -\frac{1}{l_p} \psi'+ \Pi'_0, - \frac{1}{l_p} \Big), \nonumber
\\ &=& \Big(E_p+ \frac{\varphi_0 - \psi\varphi_1 }{\gamma \Big[ 1- \psi + (\varphi_0 - \psi \varphi_1 )  \Big]}, - E_p \Big). \eea
These points are on a line $ p'_1=-E_p/c$ which has been drawn as a dashed line in the Fig. 3

\begin{figure}
  \centering
    \includegraphics[width=1.2\textwidth]{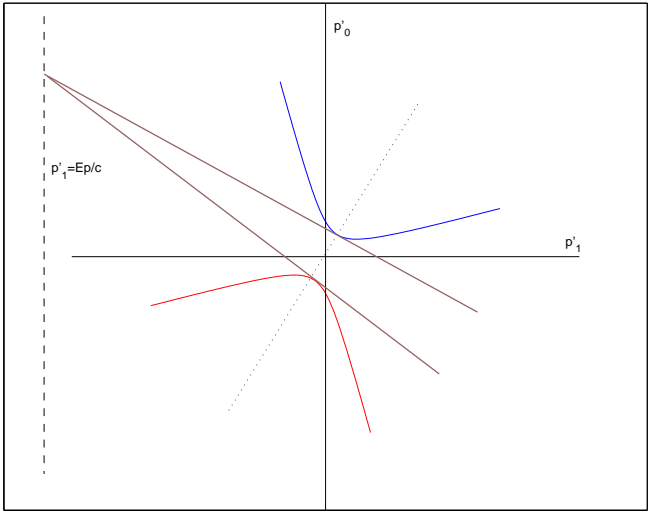}\\
  \caption{Images of the two parallel lines in the $(p'_0,p'_1)$ under the MS transformations }\label{g1}
\end{figure}

\section{Definition of the Velocity }

In the standard special relativity by taking derivative of
\be E^2 = p^2 + m^2 \ee
with respect to p, we obtain  \be \frac{dE}{dp} = \frac{p}{E},\ee
which shows that in the special relativity two expressions for velocities
 $v_H$ and $v_G $ are equal
\be  v_H=v_G.\ee

For the MS transformations, we can find a
relation between these two definitions of the velocity by taking derivative from
the dispersion relation for the MS DSR
\be \frac{E^2 - p^2}{(1 - E/E_p)}= m^2,\ee
with respect to $p$, which yields
\be \frac{dE}{dp}=\frac{p}{E \Big[ 1 + \frac{m^2}{ E E_p}(1 - E/E_p)\Big]  },\ee
in these formulas $E_p$ is the Planck energy. Thus,
\be v_H = \frac{v_G}{ 1 + \frac{m^2}{E E_p}(1 - E/E_p)}. \ee
As evidently seen from this relation if we define $v_G$ as independent parameter
from the mass or energy of the particle, then $v_H$ will depend on the mass of
the particle. Thus, we can't get ride of energy or mass dependency in the
definitions of these velocities. There is a long discussion in the DSR literature
that which formula should be interpreted as velocity $dE/dp$ or $p/E$ \cite{Ame1, Luk, Rem, Gra, Kos, Mig1, Das, Alo}.

In the MS transformations, because of the equality of the 
denominators, defining $v_G=p/E$ as velocity is like to the Lorentz
transformations case and it has a well defined rule under the MS transformations.
On the other hand , if we define the velocity as $v_H = \frac{dE}{dp} $ , this velocity will
 depend on the energy and momentum of the particle as seen from Eq.(\ref{eq13}). 

These is also an important difference between expressions for the combination rule of 
the velocities in the Lorentz transformations case Eq.(\ref{eq4}) and the MS 
transformations case Eq.(\ref{eq13}). The value of
$\Pi_0 $  , which depends on  initial point $(\varphi_0,\varphi_1) $ of
a straight line in the momentum space has not been entered in the expression for the
combination rule in the Lorentz case. But, the appearance of $\Pi_0$ in Eq.(\ref{eq13}) 
can be intercepted as a trace of an energy scale $E_p$ in the MS transformations.

Also, the meaning of a line in the momentum space or momentum plane $(p_0,p_1)$
is not so clear. We usually identify any point $(E, \vec{p})$ in the momentum
space as an energy and momentum of a physical particle. For a free particle
which has a constant energy and momentum, this point will remain stationary in this space.
 If we apply any force on this particle which can modify the energy and momentum of this particle
we will reach to another point of the momentum space.

\section{The general Linear- Fractional transformations
and MS DSR}

In the Momentum space we can write a general Linear- Fractional transformations
between$(p_0 p_1)$ and $(p'_0  p'_1)$ systems as
\be  \label{eq17} p'_0 = \frac{a_0 + A_{00}p_0+ A_{01}p_1 }{1 + \alpha_0p_0+ \alpha_1p_1 }, ~~~
 p'_1 = \frac{a_1 +   A_{10}p_0 + A_{11}p_1 }{1 + \alpha_0p_0+ \alpha_1p_1  },  \ee

in which $A_{\mu\nu} $ is the elements of a linear transformations matrix elements and $\alpha_0$ and
$ \alpha_1 $ is the parameters which shows deviations from the linearity. For the MS transformations we have

 \be \label{eq18} [A_{\mu\nu}]= \left(\begin{array}{cc} \gamma & -u\gamma \\-u\gamma & \gamma \\\end{array}\right), \ee
 \be  \label{eq19} \alpha_0 = −l_p(\gamma - 1),~~~ \alpha_1 = -l_pu\gamma.\ee
 By differentiating Eq.(\ref{eq17}) we will find $v'_H= dp'_0/dp'_1$ as

 \bea  v'_H =\frac{\Big[(A_{00} -\alpha_0a_0)v + (A_{01} - \alpha_1a_0) + (A_{01}\alpha_0 - A_{00}\alpha_1)(p_0 - vp_1) \Big]}
 {\Big[(A_{10} -\alpha_0a_1)v + (A_{11} - \alpha_1a_1) + (A_{11}\alpha_0 - A_{10}\alpha_1)(p_0 - vp_1 )\Big]} \eea
or in the compact form
\be v'_H = \frac{A_{00}v + A_{01} - p'_1(\alpha_1v + \alpha_0)}{A_{10}v + A_{11} -p'_0(\alpha_1v +\alpha_0)}. \ee
For the MS transformations case we find
\be v'_H = \frac{ v - u + l_pu(p_0 -vp_1)}{1 - uv -l_p(\gamma -1)(p_0 - vp_1)/\gamma }. \ee
If we take a line in the momentum space as $p_0 = vp_1 + \Pi_0 $ we reach to the
Eq.(\ref{eq13}), which is a special form of this equation. We had some discussions
in the end of section 2 for the energy and momentum dependency of the  $v'_H$.

\vspace{3cm}

\end{document}